\documentclass{svjour3}

\begin{document}

\title{Galactic tide in a noninertial frame of reference}
\author{J. Kla\v{c}ka}
\institute{Faculty of Mathematics,
Physics and Informatics, Comenius University \\
Mlynsk\'{a} dolina, 842 48 Bratislava, Slovak Republic \\
\email{klacka@fmph.uniba.sk}}

\date{}

\authorrunning{J. Kla\v{c}ka}
\titlerunning{Galactic tide in a noninertial frame of reference}
\maketitle

\begin{abstract}
Equation of motion and the vector of perturbing acceleration (force)
for the galactic tide in a noninertial frame of
reference is derived. The noninertial reference frame is rotating
with a fixed angular velocity $\vec{\omega}$ $=$ $-$ $\omega_{0}$
$\hat{\vec{z}}$ with respect to the inertial frame of reference of the Galaxy.
$\vec{\omega}$ is the angular velocity
of the solar rotation (rotation of the Local Standard of Rest)
around the galactic center, the unit vector $\hat{\vec{z}}$ is oriented
toward the north pole of the Galaxy: the Sun is always situated in the
plane $y'$ $=$ 0 ($x' - z'$-plane).
The equation of motion can be applied to the dynamics
of the Oort cloud of comets.
Relations for calculation of the osculating orbital elements
are presented and a new integral of motion is derived for the conventional
approach in modelling of the effect of the galactic tidal field.


\end{abstract}

\section{Introduction}
Global galactic gravitational field influences motion of a comet in the
Oort cloud in the form of the galactic tide. The effect of the galactic
tide was physicaly treated by Kla\v{c}ka (2009).

This paper presents the equation of motion in a noninertial frame of reference
$S'$ (primed quantities). The frame of reference $S'$ is rotating with a
constant angular velocity $\vec{\omega}$ $=$ $-$ $\omega_{0}$ $\hat{\vec{z}}$
around the galactic center with respect to the galactic inertial frame of
reference $S$ (it's origin is at the center of the Galaxy, $x-$ and $y-$axes
lie in the galactic equatorial plane). The unit vector $\hat{\vec{z}}$ is
oriented toward the north pole of the Galaxy, the plane $z$ $=$ $z'$ $=$ 0
is the plane of the galactic equator. The Sun is always situated in the plane
$y'$ $=$ 0 with respect to the noninertial reference frame $S'$.

Both sides (left-hand and right-hand sides) of the derived equation of motion
contain the quantities measured in the noninertial frame of reference. This is
consistent with physics and not the only difference from the conventional
approach, as it is presented by  Heisler and Tremaine (1986, Eqs. 4 and 6), or,
as for the newest papers, e.g., by Dybczynski et al. (2008).
Dybczynski et al. (2008) write on p. 347:
``The $O x' y' z'-$coordinate system rotates on the large timescale of our
simulation.'' ``In the modified heliocentric galactic $O x' y' z' -$coordinate
system, in which the $x'-$axis is orientated outward from the Galactic centre
and the $z'-$axis is orientated toward the South Galactic pole, the vector
of perturbing force can be written as $\vec{F}$ $=$
( $K_{x} x'$, $K_{y} y'$, $K_{z} z'$ ), where $x'$, $y'$, $z'$ are the
rectangular coordinates of the given TP (test particle) in the modified
heliocentric galactic-coordinate-system, and
$K_{x}$ $=$ $(A - B) (3 A + B)$,
$K_{y}$ $=$ $-~(A - B)^{2}$,
$K_{z}$ $=$ $-~[ 4 \pi k^{2} \rho_{GM} - 2 (B^{2} - A^{2})$].''
Our equation of motion contains physical terms connected with the
noninertiality of the reference frame. The corresponding part of the
equation of motion is not consistent
with the above cited description. Moreover, our equation of motion
contains new gravitational terms.

\section{Equation of motion in an inertial frame of reference}
We are interested in motion of a comet with respect to the Sun, if gravity
of the Sun and Galaxy act. Currently, the Sun is situated
$R_{0}$ $=$ 8 $kpc$ from the center of the Galaxy and
30 $pc$ above the galactic equatorial plane ($Z_{0}$ $=$ 30 $pc$).
Besides rotational motion with the speed ($A$ $-$ $B$) $R_{0}$ the Sun moves
with the speed 7.3 $km/s$ in the direction normal to the galactic plane.
Positional vector of the comet with respect to the Sun is
$\vec{r}$ $=$ ($\xi$, $\eta$, $\zeta$) in the inertial frame of reference $S$.
Equation of motion in the inertial frame of reference yields
\begin{eqnarray}\label{1}
\frac{d^{2} \xi}{dt^{2}} &=& - ~ \frac{G M_{\odot}}{r^{3}} ~ \xi
~+~ ( A - B ) \left [ A + B + 2 A \cos \left ( 2 ~ \omega_{0} t \right )
 \right ] ~ \xi
\nonumber \\
& & -~ 2 A ( A - B ) \sin \left ( 2 ~ \omega_{0} t \right ) ~\eta
\nonumber \\
& & +~ 2~ ( A - B )^{2} ~( \Gamma_{1} ~-~ \Gamma_{2} ~ Z_{0}^{2} )  ~R_{0}
~ Z_{0} ~\cos \left ( \omega_{0} t \right ) ~\zeta ~,
\nonumber \\
\frac{d^{2} \eta}{dt^{2}} &=& - ~ \frac{G M_{\odot}}{r^{3}} ~ \eta
~-~  2 A ( A - B ) \sin \left ( 2 ~ \omega_{0} t \right ) ~ \xi
\nonumber \\
& & +~ ( A - B ) \left [ A + B - 2 A \cos \left ( 2 ~ \omega_{0} t \right )
 \right ] ~ \eta
\nonumber \\
& & -~ 2~ ( A - B )^{2} ~( \Gamma_{1} ~-~ \Gamma_{2} ~ Z_{0}^{2} )  ~R_{0}
~ Z_{0} ~\sin \left ( \omega_{0} t \right ) ~\zeta ~,
\nonumber \\
\frac{d^{2} \zeta}{dt^{2}} &=& - ~ \frac{G M_{\odot}}{r^{3}} ~ \zeta
~-~ \left [ 4 ~\pi ~G ~\varrho ~+~
2 \left ( A^{2} ~-~ B^{2} \right ) \right ] ~\zeta
\nonumber \\
& & -~ 4 ~\pi ~G ~\varrho' ~
    Z_{0} \left [ \xi ~\cos \left ( \omega_{0} t \right )  ~-~
    \eta ~\sin \left ( \omega_{0} t \right )  \right ] ~,
\nonumber \\
\frac{d^{2} Z_{0}}{dt^{2}} &=& -~ \left [ 4 ~\pi ~G ~\varrho ~+~
2 \left ( A^{2} ~-~ B^{2} \right ) \right ] ~Z_{0} ~,
\nonumber \\
r &=& \sqrt{\xi ^{2} ~+~ \eta ^{2} ~+~ \zeta ^{2}} ~,
\nonumber \\
\omega_{0} &=& A ~-~ B ~,
\end{eqnarray}
where $G$ is the gravitational constant, $M_{\odot}$ is the mass
of the Sun and the numerical values of the other relevant quantities are
\begin{eqnarray}\label{2}
A &=& 14.2 ~ km ~s^{-1} ~kpc^{-1} ~,
\nonumber \\
B &=& -~ 12.4 ~ km ~s^{-1} ~kpc^{-1} ~,
\nonumber \\
\Gamma_{1} &=& 0.124 ~kpc^{-2} ~,
\nonumber \\
\Gamma_{2} &=& 1.586 ~kpc^{-4} ~,
\nonumber \\
\varrho &=& 0.130 ~M_{\odot}~ pc^{-3} ~,
\nonumber \\
\varrho' &=& -~ 0.037 ~M_{\odot} ~pc^{-3} ~kpc^{-1} ~,
\end{eqnarray}
see Eqs. (26)-(27) in Kla\v{c}ka (2009).
If one wants to use other values of the Oort constants $A$ and $B$, then he can
use the following equation for calculation of mass density in the neighborhood
of the Sun:
\begin{eqnarray}\label{3}
\varrho &=&  \varrho_{disk} ~+~ \varrho_{halo} ~,
\nonumber \\
\varrho_{disk} &=& 0.126 ~ M_{\odot} ~ pc^{-3} ~,
\nonumber \\
\varrho_{halo} &=& (4 \pi G)^{-1} [ X(Galaxy) + X(disk) + X(bulge) ]  ~,
\nonumber \\
X(Galaxy) &\equiv& - (A - B) \times (A + 3 B)
\nonumber \\
X(disk) &=&  -~396.90 ~km^{2} ~s^{-2} ~kpc^{-2} ~,
\nonumber \\
X(bulge) &=& -~0.65 ~km^{2} ~s^{-2} ~kpc^{-2} ~.
\end{eqnarray}
Eqs. (22) of Kla\v{c}ka (2009) can be used.

\subsection{Vector of perturbing acceleration}
On the basis of Eqs. (1) we can state that dominant force is represented
by gravity of the Sun and the perturbing force is given by gravity of the
Galaxy. The perturbing acceleration due to the action of the Galaxy is
\begin{eqnarray}\label{4}
\vec{F} &=& F_{x} ~\hat{\vec{x}} ~+~
			 F_{y} ~\hat{\vec{y}} ~+~ F_{z} ~\hat{\vec{z}} ~,
\nonumber \\
F_{x} &=& ( A - B ) \left [ A + B + 2 A \cos \left ( 2 ~ \omega_{0} t \right )
 \right ] ~ \xi
\nonumber \\
& & -~ 2 A ( A - B ) \sin \left ( 2 ~ \omega_{0} t \right ) ~\eta
\nonumber \\
& & +~ 2~ ( A - B )^{2} ~( \Gamma_{1} ~-~ \Gamma_{2} ~ Z_{0}^{2} )  ~R_{0}
~ Z_{0} ~\cos \left ( \omega_{0} t \right ) ~\zeta ~,
\nonumber \\
F_{y} &=& -~ 2 A ( A - B ) \sin \left ( 2 ~ \omega_{0} t \right ) ~ \xi
\nonumber \\
& & +~ ( A - B ) \left [ A + B - 2 A \cos \left ( 2 ~ \omega_{0} t \right )
 \right ] ~ \eta
\nonumber \\
& & -~ 2~ ( A - B )^{2} ~( \Gamma_{1} ~-~ \Gamma_{2} ~ Z_{0}^{2} )  ~R_{0}
~ Z_{0} ~\sin \left ( \omega_{0} t \right ) ~\zeta ~,
\nonumber \\
F_{z} &=& -~ \left [ 4 ~\pi ~G ~\varrho ~+~
2 \left ( A^{2} ~-~ B^{2} \right ) \right ] ~\zeta
\nonumber \\
& & -~ 4 ~\pi ~G ~\varrho' ~
    Z_{0} \left [ \xi ~\cos \left ( \omega_{0} t \right )  ~-~
    \eta ~\sin \left ( \omega_{0} t \right )  \right ] ~.
\end{eqnarray}

The perturbing acceleration ceases to exist when
$A$ $\equiv$ $B$ $\equiv$ $\varrho$ $\equiv$ 0.

\section{Equation of motion in a noninertial frame of reference}
We are interested in the motion of a comet with respect to the Sun
in the noninertial frame of reference $S'$. The noninertial frame of reference
is defined as the frame rotating with a constant angular velocity
$\vec{\omega}$ $=$ $\omega$ $\hat{\vec{z}}$
around the galactic center with respect to the galactic inertial frame of
reference $S$ (it's origin is at the center of the Galaxy, $x-$ and $y-$axes
lie in the galactic equatorial plane). The unit vector $\hat{\vec{z}}$ is
oriented toward the north pole of the Galaxy, the plane $z$ $=$ $z'$ $=$ 0
is the plane of the galactic equator. The Sun is moves in the plane
$y'$ $=$ 0 with respect to the noninertial reference frame $S'$. The Sun's
galactocentric position vector is
\begin{eqnarray}\label{5}
\vec{R}_{0} &=& R_{0}~ \cos ( - ~\omega_{0} ~t ) ~\hat{\vec{x}} ~+~
R_{0}~ \sin ( - ~\omega_{0} ~t ) ~\hat{\vec{y}} ~+~
Z_{0}~\hat{\vec{z}} ~,
\end{eqnarray}
in the inertial frame of reference and
\begin{eqnarray}\label{6}
\vec{R}_{0} ' &=& R_{0}~ \hat{\vec{x}}' ~+~
Z_{0}~\hat{\vec{z}}' ~,
\end{eqnarray}
in the noninertial frame of reference, since $\vec{R}_{0} '$ $=$ $\vec{R}_{0}$.
The sign minus at $\omega_{0}$
denotes negative (clockwise) orientation/direction of the motion of the Sun,
$\vec{\omega}$ $=$ $\omega$ $\hat{\vec{z}}$ $=$
$-$ $\omega_{0}$ $\hat{\vec{z}}$,
$R_{0}$ $=$ 8 $kpc$.

\subsection{Transformations}
The unit vectors $\hat{\vec{x}}$, $\hat{\vec{y}}$ and
$\hat{\vec{z}}$ form an orthonormal basis and the right-handed system in the
inertial frame of reference $S$. The same holds for the primed unit vectors
of the rotating noninertial system $S'$, and
\begin{eqnarray}\label{7}
\hat{\vec{x}}' &=& \cos ( \omega ~t ) ~\hat{\vec{x}} ~+~
		   \sin ( \omega ~t ) ~\hat{\vec{y}} ~,
\nonumber \\
\hat{\vec{y}}' &=& -~ \sin ( \omega ~t ) ~\hat{\vec{x}} ~+~
		   \cos ( \omega ~t ) ~\hat{\vec{y}} ~,
\nonumber \\
\hat{\vec{z}}' &=& \hat{\vec{z}}  ~.
\end{eqnarray}
The inverse relations are
\begin{eqnarray}\label{8}
\hat{\vec{x}} &=& \cos ( \omega ~t ) ~\hat{\vec{x}} ' ~-~
		  \sin ( \omega ~t ) ~\hat{\vec{y}} ' ~,
\nonumber \\
\hat{\vec{y}} &=& \sin ( \omega ~t ) ~\hat{\vec{x}} ' ~+~
		  \cos ( \omega ~t ) ~\hat{\vec{y}} ' ~,
\nonumber \\
\hat{\vec{z}} &=& \hat{\vec{z}} ' ~.
\end{eqnarray}

Let us consider a point mass with position vector $\vec{r}$ $=$ $\vec{r} '$.
We have $\vec{r}$ $=$ ($x$, $y$, $z$) and $\vec{r}'$ $=$ ($x'$, $y'$, $z'$),
or,
$\vec{r}$ $=$ $x$ $\hat{\vec{x}}$ $+$ $y$ $\hat{\vec{y}}$ $+$ $z$ $\hat{\vec{z}}$,
$\vec{r} '$ $=$ $x '$ $\hat{\vec{x}} '$ $+$ $y '$ $\hat{\vec{y}} '$ $+$
$z '$ $\hat{\vec{z}} '$.
The components in the inertial and noninertial
reference frames are related through the relations
\begin{eqnarray}\label{9}
x' &=& x~ \cos ( \omega ~t ) ~+~
       y~ \sin ( \omega ~t ) ~,
\nonumber \\
y' &=& -~ x~ \sin ( \omega ~t ) ~+~
	  y~ \cos ( \omega ~t ) ~,
\nonumber \\
z' &=& z  ~.
\end{eqnarray}
The orthogonal transformation given by Eqs. (9) immediately offers
the inverse transformation:
\begin{eqnarray}\label{10}
x &=& x'~ \cos ( \omega ~t )  ~-~
      y'~ \sin ( \omega ~t ) ~,
\nonumber \\
y &=& x'~ \sin ( \omega ~t ) ~+~
      y'~ \cos ( \omega ~t ) ~,
\nonumber \\
z &=& z'  ~.
\end{eqnarray}
The velocity with respect to the system $S'$ is $\vec{v}'$ $\equiv$
$d' \vec{r} ' / dt$ $\equiv$ ($dx'/dt$) $\hat{\vec{x}}'$ $+$
($dy'/dt$) $\hat{\vec{y}}'$ $+$ ($dz'/dt$) $\hat{\vec{z}}'$.
The relation $\vec{v}'$ $=$ ($dx'/dt$) $\hat{\vec{x}}'$ $+$
($dy'/dt$) $\hat{\vec{y}}'$ $+$ ($dz'/dt$) $\hat{\vec{z}}'$ and Eqs. (8) yield
\begin{eqnarray}\label{11}
( \vec{v}' )_{x} &\equiv& \vec{v}' \cdot \hat{\vec{x}} =
       \frac{d x '}{dt} ~ \cos ( \omega ~t ) ~-~
       \frac{d y '}{dt} ~ \sin ( \omega ~t ) ~,
\nonumber \\
( \vec{v}' )_{y} &\equiv& \vec{v}' \cdot \hat{\vec{y}} =
       \frac{d x '}{dt} ~ \sin ( \omega ~t ) ~+~
       \frac{d y '}{dt} ~ \cos ( \omega ~t ) ~,
\nonumber \\
( \vec{v}' )_{z} &\equiv& \vec{v}' \cdot \hat{\vec{z}} =
\frac{d z '}{dt} ~.
\end{eqnarray}
It can be said that Eqs. (11) hold on the basis of vectorial transformation
defined by Eqs. (10), because any vector in the rotating frame must
project onto $x-$, $y-$ and $z-$ axes in the same way as any other vector
(Kittel et al. 1965, p. 85). As for the acceleration vector, we have
$\vec{a}'$ $\equiv$ $d' \vec{v}' / dt$ $=$ $d'$ $^{2} \vec{r}' / dt^{2}$
with respect to the system $S'$, and
\begin{eqnarray}\label{12}
( \vec{a}' )_{x} &\equiv& \vec{a}' \cdot \hat{\vec{x}} =
       \frac{d^{2} x '}{dt^{2}} ~ \cos ( \omega ~t ) ~-~
       \frac{d^{2} y '}{dt^{2}} ~ \sin ( \omega ~t ) ~,
\nonumber \\
( \vec{a}' )_{y} &\equiv& \vec{a}' \cdot \hat{\vec{y}} =
       \frac{d^{2} x '}{dt^{2}} ~ \sin ( \omega ~t ) ~+~
       \frac{d^{2} y '}{dt^{2}} ~ \cos ( \omega ~t ) ~,
\nonumber \\
( \vec{a}' )_{z} &\equiv& \vec{a}' \cdot \hat{\vec{z}} =
\frac{d^{2} z '}{dt^{2}} ~.
\end{eqnarray}
Similarly,
\begin{eqnarray}\label{13}
( \vec{\omega} \times \vec{v}' )_{x} &=& -~\omega ~( \vec{v}' )_{y} =
       -~\omega ~ \left \{
       \frac{d x '}{dt} ~ \sin ( \omega ~t ) ~+~
       \frac{d y '}{dt} ~ \cos ( \omega ~t ) \right \} ~,
\nonumber \\
( \vec{\omega} \times \vec{v}' )_{y} &=& +~\omega ~( \vec{v}' )_{x} =
       +~ \omega ~ \left \{
       \frac{d x '}{dt} ~ \cos ( \omega ~t ) ~-~
       \frac{d y '}{dt} ~ \sin ( \omega ~t ) \right \}
\end{eqnarray}
and
\begin{eqnarray}\label{14}
[ \vec{\omega} \times ( \vec{\omega} \times \vec{r}' ) ]_{x} &=&
       -~\omega ^{2} ~( \vec{r}' )_{x} = -~\omega ^{2} ~ \left \{
       x ' \cos ( \omega ~t ) - y ' \sin ( \omega ~t ) \right \} ~,
\nonumber \\
\left [ \vec{\omega} \times ( \vec{\omega} \times \vec{r}' ) \right ]_{y} &=&
       -~\omega ^{2} ~( \vec{r}' )_{y} = -~ \omega ^{2} ~ \left \{
       x ' \sin ( \omega ~t ) + y ' \cos ( \omega ~t ) \right \} ~.
\end{eqnarray}

On the basis of Eqs. (7)-(14) we can write
\begin{eqnarray}\label{15}
\vec{a} &=& \vec{a}' ~+~ 2~ \vec{\omega} \times \vec{v}'  ~+~
 \vec{\omega} \times ( \vec{\omega} \times \vec{r}' ) ~,
\end{eqnarray}
for accelerations in the inertial (unprimed quantities) and
noninertial (primed quantities) reference frames
(see,  e.g., Kittel et al. 1965, pp. 83-85).

\subsection{Galactic tide in the noninertial frame of reference}
On the basis of Eqs. (10)-(15), we can write for the $x-$ ($\xi-$) component of
the acceleration given by Eqs. (1):
\begin{eqnarray}\label{16}
LHS_{x} &=& RHS_{x} ~,
\nonumber \\
LHS_{x} &\equiv&
- ~ \frac{G M_{\odot}}{r^{3}} ~ \left [ \xi' \cos \left ( \omega_{0} t \right )
~+~ \eta ' ~\sin \left ( \omega_{0} t \right ) \right ]
\nonumber \\
& & +~( A - B ) \left [ ( A + B ) + 2 A \cos \left ( 2 \omega_{0} t \right ) \right ]
 \left [ \xi' \cos \left ( \omega_{0} t \right )
~+~ \eta ' ~\sin \left ( \omega_{0} t \right ) \right ]
\nonumber \\
& & -~ 2 A ( A - B ) ~\sin \left ( 2 \omega_{0} t \right )
 \left [ -~ \xi' \sin \left ( \omega_{0} t \right )
~+~ \eta ' ~\cos \left ( \omega_{0} t \right ) \right ]
\nonumber \\
& & +~ 2~ ( A - B )^{2} ~( \Gamma_{1} ~-~ \Gamma_{2} ~ Z_{0}^{2} )  ~R_{0}
~ Z_{0} ~\cos \left ( \omega_{0} t \right ) ~\zeta '~,
\nonumber \\
RHS_{x} &\equiv& \frac{d^{2} \xi'}{dt^{2}} ~\cos \left ( \omega_{0} t \right ) ~+~
\frac{d^{2} \eta'}{dt^{2}} ~\sin \left ( \omega_{0} t \right )
\nonumber \\
& & +~ 2 \omega_{0} \left [ -~ \frac{d\xi'}{dt}~
\sin \left ( \omega_{0} t \right )
~+~ \frac{d \eta '}{dt} ~\cos \left ( \omega_{0} t \right ) \right ]
\nonumber \\
& & -~ \omega_{0}^{2} \left [ \xi' ~ \cos \left ( \omega_{0} t \right )
+ \eta ' ~\sin \left ( \omega_{0} t \right ) \right ]
\end{eqnarray}
or, after some algebra,
\begin{eqnarray}\label{17}
LHS_{x} &=& RHS_{x} ~,
\nonumber \\
LHS_{x} &\equiv&
- ~ \frac{G M_{\odot}}{r^{3}} ~ \left [ \xi' \cos \left ( \omega_{0} t \right )
~+~ \eta ' ~\sin \left ( \omega_{0} t \right ) \right ]
\nonumber \\
& & +~( A - B ) \left [ ( 3 A + B ) ~
 \xi' \cos \left ( \omega_{0} t \right ) ~-~ ( A - B ) ~
 \eta ' ~\sin \left ( \omega_{0} t \right ) \right ]
\nonumber \\
& & +~ 2~ ( A - B )^{2} ~( \Gamma_{1} ~-~ \Gamma_{2} ~ Z_{0}^{2} )  ~R_{0}
~ Z_{0} ~\cos \left ( \omega_{0} t \right ) ~\zeta ' ~,
\nonumber \\
RHS_{x} &\equiv& \frac{d^{2} \xi'}{dt^{2}} ~\cos \left ( \omega_{0} t \right ) ~+~
\frac{d^{2} \eta'}{dt^{2}} ~\sin \left ( \omega_{0} t \right )
\nonumber \\
& & +~ 2 \omega_{0} \left [ -~ \frac{d\xi'}{dt}~
\sin \left ( \omega_{0} t \right )
~+~ \frac{d \eta '}{dt} ~\cos \left ( \omega_{0} t \right ) \right ]
\nonumber \\
& & -~ \omega_{0}^{2} \left [ \xi' ~ \cos \left ( \omega_{0} t \right )
+ \eta ' ~\sin \left ( \omega_{0} t \right ) \right ] ~.
\end{eqnarray}

On the basis of Eqs. (10)-(15), we can write for the $y-$ ($\eta-$) component
of the acceleration given by Eqs. (1):
\begin{eqnarray}\label{18}
LHS_{y} &=& RHS_{y} ~,
\nonumber \\
LHS_{y} &\equiv&
- ~ \frac{G M_{\odot}}{r^{3}} ~ \left [ -~\xi' \sin \left ( \omega_{0} t \right )
~+~ \eta ' ~\cos \left ( \omega_{0} t \right ) \right ]
\nonumber \\
& & -~2 A ( A - B ) ~ \sin \left ( 2 \omega_{0} t \right )
 \left [ \xi' \cos \left ( \omega_{0} t \right )
~+~ \eta ' ~\sin \left ( \omega_{0} t \right ) \right ]
\nonumber \\
& & +~ ( A - B ) [ A + B - 2 A ~\cos \left ( 2 \omega_{0} t \right ) ]
 \left [ -~ \xi' \sin \left ( \omega_{0} t \right )
~+~ \eta ' ~\cos \left ( \omega_{0} t \right ) \right ]
\nonumber \\
& & -~ 2~ ( A - B )^{2} ~( \Gamma_{1} ~-~ \Gamma_{2} ~ Z_{0}^{2} )  ~R_{0}
~ Z_{0} ~\sin \left ( \omega_{0} t \right ) ~\zeta ' ~,
\nonumber \\
RHS_{y} &\equiv& -~\frac{d^{2} \xi'}{dt^{2}} ~\sin \left ( \omega_{0} t \right ) ~+~
\frac{d^{2} \eta'}{dt^{2}} ~\cos \left ( \omega_{0} t \right )
\nonumber \\
& & -~ 2 \omega_{0} \left [ \frac{d\xi'}{dt}~
\cos \left ( \omega_{0} t \right )
~+~ \frac{d \eta '}{dt} ~\sin \left ( \omega_{0} t \right ) \right ]
\nonumber \\
& & -~ \omega_{0}^{2} \left [ -~\xi' ~ \sin \left ( \omega_{0} t \right )
+ \eta ' ~\cos \left ( \omega_{0} t \right ) \right ]
\end{eqnarray}
or, after some algebra,
\begin{eqnarray}\label{19}
LHS_{y} &=& RHS_{y} ~,
\nonumber \\
LHS_{y} &\equiv&
- ~ \frac{G M_{\odot}}{r^{3}} ~ \left [ -~\xi' \sin \left ( \omega_{0} t \right )
~+~ \eta ' ~\cos \left ( \omega_{0} t \right ) \right ]
\nonumber \\
& & -~( A - B ) \left [  ( 3 A + B ) ~
 \xi' \sin \left ( \omega_{0} t \right ) ~+~ ( A - B ) ~
 \eta ' ~\cos \left ( \omega_{0} t \right ) \right ]
\nonumber \\
& & -~ 2~ ( A - B )^{2} ~( \Gamma_{1} ~-~ \Gamma_{2} ~ Z_{0}^{2} )  ~R_{0}
~ Z_{0} ~\sin \left ( \omega_{0} t \right ) ~\zeta ' ~,
\nonumber \\
RHS_{y} &\equiv& -~\frac{d^{2} \xi'}{dt^{2}} ~\sin \left ( \omega_{0} t \right ) ~+~
\frac{d^{2} \eta'}{dt^{2}} ~\cos \left ( \omega_{0} t \right )
\nonumber \\
& & -~ 2 \omega_{0} \left [ \frac{d\xi'}{dt}~
\cos \left ( \omega_{0} t \right )
~+~ \frac{d \eta '}{dt} ~\sin \left ( \omega_{0} t \right ) \right ]
\nonumber \\
& & -~ \omega_{0}^{2} \left [ -~\xi' ~ \sin \left ( \omega_{0} t \right )
+ \eta ' ~\cos \left ( \omega_{0} t \right ) \right ] ~.
\end{eqnarray}

Eqs. (17) and (19) yield
\begin{eqnarray}\label{20}
\frac{d^{2} \xi'}{dt^{2}} &=& - ~ \frac{G M_{\odot}}{r^{3}} ~\xi'
~+~( A - B ) ( 3 A + B ) ~  \xi'
\nonumber \\
& & +~ 2~ ( A - B )^{2} ~( \Gamma_{1} ~-~ \Gamma_{2} ~ Z_{0}^{2} )  ~R_{0}
~ Z_{0} ~\zeta '
\nonumber \\
& & -~ 2 \omega_{0} ~ \frac{d\eta'}{dt} ~+~ \omega_{0}^{2} ~\xi'  ~,
\nonumber \\
\frac{d^{2} \eta'}{dt^{2}} &=& - ~ \frac{G M_{\odot}}{r^{3}} ~\eta'  ~-~
( A - B )^{2}~	\eta' ~+~
 2 \omega_{0} ~ \frac{d\xi'}{dt} ~+~ \omega_{0}^{2} ~\eta'  ~.
\end{eqnarray}

Finally, Eqs. (1) and (20) yield
\begin{eqnarray}\label{21}
\frac{d^{2} \xi'}{dt^{2}} &=& - ~ \frac{G M_{\odot}}{r^{3}} ~\xi'
			      ~+~ 4~ A~ ( A ~-~ B ) ~ \xi'
\nonumber \\
& & +~ 2~ ( A - B )^{2} ~( \Gamma_{1} ~-~ \Gamma_{2} ~ Z_{0}^{2} )  ~R_{0}
~ Z_{0} ~\zeta ' ~-~ 2~ ( A ~-~ B ) ~ \frac{d\eta'}{dt}  ~,
\nonumber \\
\frac{d^{2} \eta'}{dt^{2}} &=& - ~ \frac{G M_{\odot}}{r^{3}} ~\eta'  ~+~
 2~ ( A ~-~ B ) ~ \frac{d\xi'}{dt} ~,
\nonumber \\
\frac{d^{2} \zeta'}{dt^{2}} &=& - ~ \frac{G M_{\odot}}{r^{3}} ~ \zeta '
~-~ \left [ 4 ~\pi ~G ~\varrho ~+~
2~ \left ( A^{2} ~-~ B^{2} \right ) \right ] ~\zeta '
\nonumber \\
& & -~ 4 ~\pi ~G ~\varrho' ~ Z_{0}  ~ \xi' ~,
\nonumber \\
\frac{d^{2} Z_{0}}{dt^{2}} &=& -~ \left [ 4 ~\pi ~G ~\varrho ~+~
2 ~\left ( A^{2} ~-~ B^{2} \right ) \right ] ~Z_{0} ~,
\nonumber \\
r &=& \sqrt{\xi^{'~2} ~+~ \eta^{'~2} ~+~ \zeta^{'~2}} ~.
\end{eqnarray}

\subsection{Vector of perturbing acceleration}
The perturbing force is given by the effect of Galaxy, according to Sec. 2.1.
This effect is represented by the Oort constants $A$, $B$ and mass density
$\varrho$.

The vector of the perturbing acceleration acting on the comet is
\begin{eqnarray}\label{22}
\vec{F}' &=& ( \vec{F}' )_{\xi '} ~\hat{\vec{\xi}} ' ~+~
( \vec{F}' )_{\eta '} ~\hat{\vec{\eta}} ' ~+~
( \vec{F}' )_{\zeta '} ~\hat{\vec{\zeta}} ' ~,
\nonumber \\
( \vec{F}' )_{\xi '} &=& 2~ ( A ~-~ B ) ~\left \{ 2~ A~  \xi'
~+~ ( A - B )~( \Gamma_{1} ~-~ \Gamma_{2} ~ Z_{0}^{2} )  ~R_{0}
~ Z_{0} ~\zeta ' ~-~ \frac{d\eta'}{dt} \right \}  ~,
\nonumber \\
( \vec{F}' )_{\eta '}  &=&  2~ ( A ~-~ B ) ~ \frac{d\xi'}{dt} ~,
\nonumber \\
( \vec{F}' )_{\zeta '} &=& -~ \left [ 4 ~\pi ~G ~\varrho ~+~
2~ \left ( A^{2} ~-~ B^{2} \right ) \right ] ~\zeta' ~-~
4 ~\pi ~G ~\varrho' ~ Z_{0}  ~ \xi' ~,
\end{eqnarray}
if Eqs. (21) are taken into account.

The effect of Galaxy is turned off when
$A$ $=$ $B$ $=$ $\varrho$ $=$ $\varrho'$ $=$ 0.
$\vec{F}'$ $=$ 0 and the two-body problem exists in this case.
This is consistent with Sec. 2.1.

\section{Discussion}
We want to concentrate on obtaining the evolution of
osculating orbital elements on the basis of solution of Eqs. (21).

\subsection{Osculating orbital elements}
Eqs. (21) offer the values of coordinates $\xi'$, $\eta'$, $\zeta'$ and
velocities $d\xi' / dt$, $d\eta' / dt$ and $d\zeta' / dt$ for a time $t$.
We have to obtain the values in the inertial frame of reference $S$.
We need $\xi$, $\eta$, $\zeta$, $d\xi / dt$, $d\eta / dt$ and $d\zeta / dt$
for the time $t$.

On the basis of Eq. (20) we obtain
\begin{eqnarray}\label{23}
\xi &=& \xi'~ \cos ( \omega_{0} ~t )  ~+~ \eta'~ \sin ( \omega_{0} ~t ) ~,
\nonumber \\
\eta &=& -~ \xi'~ \sin ( \omega_{0}~t ) ~+~ \eta'~ \cos ( \omega_{0} ~t ) ~,
\nonumber \\
\zeta &=& \zeta'  ~.
\end{eqnarray}
The velocity components $d\xi / dt$, $d\eta / dt$ and $d\zeta / dt$ can be
obtained from Eq. (23):
\begin{eqnarray}\label{24}
\frac{d \xi}{dt} &=&
       \frac{d \xi '}{dt} ~ \cos ( \omega_{0} ~t ) ~+~
       \frac{d \eta '}{dt} ~ \sin ( \omega_{0} ~t )
\nonumber \\
& & -~ \omega_{0} ~ \xi '~ \sin ( \omega_{0} ~t ) ~+~
       \omega_{0} ~ \eta '~ \cos ( \omega_{0} ~t ) ~,
\nonumber \\
\frac{d \eta}{dt} &=&
       -~ \frac{d \xi '}{dt} ~ \sin ( \omega_{0} ~t ) ~+~
       \frac{d \eta '}{dt} ~ \cos ( \omega_{0} ~t )
\nonumber \\
& & -~ \omega_{0} ~ \xi '~ \cos ( \omega_{0} ~t ) ~-~
       \omega_{0} ~ \eta '~ \sin ( \omega_{0} ~t ) ~,
\nonumber \\
\frac{d \zeta}{dt} &=& \frac{d \zeta '}{dt} ~.
\end{eqnarray}
Eqs. (24) are consistent with the relation
\begin{eqnarray}\label{25}
\vec{v} &=& \vec{v}' ~+~ \vec{\omega} \times \vec{r}'  ~,
\end{eqnarray}
since $\vec{\omega} \times \vec{r}'$ $=$ $\omega \hat{\vec{z}} '$ $\times$
($x '$ $\hat{\vec{x}} '$ $+$ $y '$ $\hat{\vec{y}} '$) $=$ $\omega$
($x '$ $\hat{\vec{y}} '$ $-$ $y '$ $\hat{\vec{x}} '$) and
( $\vec{\omega} \times \vec{r}'$ ) $\cdot$ $\hat{\vec{x}}$ $=$ $\omega$
($x '$ $\hat{\vec{y}} '$ $\cdot$ $\hat{\vec{x}}$ $-$
$y '$ $\hat{\vec{x}} '$ $\cdot$ $\hat{\vec{x}}$ ) and Eqs. (7) hold;
similar consideration can be used also for
( $\vec{\omega} \times \vec{r}'$ ) $\cdot$ $\hat{\vec{y}}$.

The evolution of the orbital elements is obtained on the basis of
Eqs. (21), (23)-(24)
and the relations presented by Kla\v{c}ka (2004) in his Eqs. (47)
(the right-hand side of the last equation in Eqs. 47 must
contain $1/e$ instead of 1). We can summarize the equations in the
following form [osculating orbital elements: $a$ -- semi-major axis;
$e$ -- eccentricity; $i$ -- inclination of the orbital plane to the
reference plane -- galactic equatorial plane;
$\Omega$ -- longitude of the ascending node; $\omega$ --
the argument of pericenter/perihelion; $\Theta$
is the position angle of the particle on the orbit, when measured
from the ascending node in the direction of the particle's motion,
$\Theta = \omega + f$]:
\begin{eqnarray}\label{26}
\vec{r} &=& ( \xi, \eta, \zeta ) ~, ~~ r = | \vec{r} | ~,
\nonumber \\
\vec{v} &\equiv& \frac{d \vec{r}}{dt} =
( \frac{d \xi}{dt}, \frac{d \eta}{dt}, \frac{d \zeta}{dt} ) ~,
\nonumber \\
E &=& \frac{1}{2} ~\vec{v}^{2} ~-~ \frac{G ~M_{\odot}}{r} ~,
\nonumber \\
\vec{H} &=& \vec{r} \times \vec{v} ~,
\nonumber \\
p &=& \frac{\vec{H}^{2}}{G ~M_{\odot}} ~,
\nonumber \\
e &=& \sqrt{1 ~+~ 2 ~\frac{p~ E}{G ~M_{\odot}}} ~,
\nonumber \\
a &=& \frac{p}{1 - e^{2}} ~,
\nonumber \\
E &=& -~ \frac{G~ M_{\odot}}{2~p} ~ \left ( 1 - e^{2} \right ) ~,
\nonumber \\
q &=& a ( 1 - e ) ~,
\nonumber \\
i &=& \arccos{ \left ( \frac{H_{z}}{| \vec{H} |} \right ) } ~,
\nonumber \\
\sin \Omega &=& \frac{H_{\xi}}{ \sqrt{H_{\xi}^{2} + H_{\eta}^{2}} } ~,
\nonumber \\
\cos \Omega &=& -~ \frac{H_{\eta}}{ \sqrt{H_{\xi}^{2} + H_{\eta}^{2}} } ~,
\nonumber \\
\sin \omega &=& \frac{| \vec{H} |}{ \sqrt{H_{\xi}^{2} + H_{\eta}^{2}} } ~
		\frac{1}{r ~e} ~\times~ S1
\nonumber \\
S1 &=& -~ \frac{\eta~H_{\xi} ~-~ \xi~H_{\eta}}{| \vec{H} |} ~
       \frac{\vec{v} \cdot \vec{e}_{R}}{\sqrt{G ~M_{\odot} / p}} ~+~
     \zeta ~\left ( \frac{\vec{v} \cdot \vec{e}_{T}}{\sqrt{G ~M_{\odot} / p}}
     ~-~ 1 \right ) ~,
\nonumber \\
\cos \omega &=& \frac{| \vec{H} |}{ \sqrt{H_{\xi}^{2} + H_{\eta}^{2}} } ~
		\frac{1}{r ~e} ~\times~ C1
\nonumber \\
C1 &=&	\frac{\eta~H_{\xi} ~-~ \xi~H_{\eta}}{| \vec{H} |} ~
	\left ( \frac{\vec{v} \cdot \vec{e}_{T}}{\sqrt{G ~M_{\odot} / p}}
	~-~ 1 \right ) ~+~ \zeta ~
       \frac{\vec{v} \cdot \vec{e}_{R}}{\sqrt{G ~M_{\odot} / p}} ~,
\nonumber \\
\vec{e}_{R} &=& \frac{\vec{r}}{r} ~,
\nonumber \\
\vec{e}_{N} &=& \frac{\vec{H}}{| \vec{H} |} ~,
\nonumber \\
\vec{e}_{T} &=& \vec{e}_{N} \times \vec{e}_{R} ~.
\end{eqnarray}

\subsection{Simple consequence of Eqs. (21)}
Eqs. (21) yield
\begin{eqnarray}\label{27}
LHS_{S'} &=& RHS_{S'} ~,
\nonumber \\
LHS_{S'} &=& \frac{d}{dt} ~\left \{ \frac{1}{2} ~ \left [
\left ( \frac{d \xi '}{dt} \right )^{2} ~+~
\left ( \frac{d \eta '}{dt} \right )^{2} ~+~
\left ( \frac{d \zeta '}{dt} \right )^{2} \right ] ~-~
\frac{G ~M_{\odot}}{r} \right .
\nonumber \\
& & \left .  -~ 2 ~A ~( A ~-~ B ) ~ ( \xi ' )^{2}
~+~ ( 2 ~\pi ~G ~\varrho ~+~
A^{2} ~-~ B^{2} ) ~ ( \zeta'  )^{2}\right \} ~,
\nonumber \\
RHS_{S'} &=&
2~ ( A - B )^{2} ~( \Gamma_{1} ~-~ \Gamma_{2} ~ Z_{0}^{2} )  ~R_{0}
~ Z_{0} ~\zeta '~\frac{d\xi'}{dt}
~-~ 4 ~\pi ~G ~\varrho' ~ Z_{0}  ~ \xi' ~\frac{d \zeta '}{dt} ~.
\end{eqnarray}

Taking into account
\begin{eqnarray}\label{28}
\vec{v}' &=& \vec{v} ~-~ \vec{\omega} \times \vec{r}  ~,
\end{eqnarray}
(Eq. 25 and the relation $\vec{r}'$ $=$ $\vec{r}$ hold), we have
\begin{eqnarray}\label{29}
T' &=& T ~-~ \vec{\omega} \cdot ( \vec{r} \times \vec{v} ) ~+~
\frac{1}{2} \left ( \vec{\omega} \times \vec{r} \right )^{2} ~,
\nonumber \\
T' &\equiv& \frac{1}{2} ~ \left [
\left ( \frac{d \xi '}{dt} \right )^{2} ~+~
\left ( \frac{d \eta '}{dt} \right )^{2} ~+~
\left ( \frac{d \zeta '}{dt} \right )^{2} \right ] ~,
\nonumber \\
T &\equiv& \frac{1}{2} ~ \left [
\left ( \frac{d \xi}{dt} \right )^{2} ~+~
\left ( \frac{d \eta}{dt} \right )^{2} ~+~
\left ( \frac{d \zeta}{dt} \right )^{2} \right ] ~,
\end{eqnarray}
where $T'$ is the kinetic energy in the rotating noninertial frame of reference
and $T$ is the kinetic energy in the inertial frame of reference. Using
\begin{eqnarray}\label{30}
\vec{\omega} &=& -~\omega_{0} ~\hat{\vec{z}} ~,
\nonumber \\
\hat{\vec{z}} \cdot ( \vec{r} \times \vec{v} ) &=&
\hat{\vec{z}} \cdot \vec{H} \equiv H_{z} ~,
\nonumber \\
H_{z} &=& \sqrt{G ~M_{\odot}~p} ~\cos i ~,
\end{eqnarray}
Eqs. (26), (27), (29) and (30) yield
\begin{eqnarray}\label{31}
LHS_{S} &=& RHS_{S} ~,
\nonumber \\
LHS_{S} &=& \frac{d}{dt} ~\left \{ -~\frac{G~M_{\odot}}{2 ~p} ~\left (
1 ~-~ e^{2} \right ) ~+~ ( A ~-~ B ) ~ \sqrt{G ~M_{\odot}~p} ~\cos i
\right .
\nonumber \\
& & \left .  + ~\frac{1}{2} ~ ( A ~-~ B )^{2} ~ \left ( \xi ^{2} ~+~
\eta ^{2} \right ) \right .
\nonumber \\
& & \left .  -~ 2 ~A ~ ( A ~-~ B ) ~ [ \xi~ \cos ( \omega_{0} ~t ) ~-~
	     \eta~ \sin ( \omega_{0} ~t )  ] ^{2} \right .
\nonumber \\
& & \left . +~ ( 2 ~\pi ~G ~\varrho ~+~
A^{2} ~-~ B^{2} ) ~ \zeta ^{2} \right \} ~,
\nonumber \\
RHS_{S} &=& 2~ ( A - B )^{2} ~( \Gamma_{1} ~-~ \Gamma_{2} ~ Z_{0}^{2} )  ~R_{0}
~ Z_{0} ~\zeta ~ \times V
\nonumber \\
& & -~ 4 ~\pi ~G ~\varrho' ~ Z_{0}  ~
[ \xi~ \cos ( \omega_{0} ~t ) ~-~ \eta~ \sin ( \omega_{0} ~t ) ]
~\frac{d \zeta}{dt}  ~,
\nonumber \\
V &=& \frac{d \xi}{dt} ~ \cos ( \omega_{0} ~t ) ~-~
       \frac{d \eta}{dt} ~ \sin ( \omega_{0} ~t )
\nonumber \\
& & -~ \omega_{0} ~ \xi ~ \sin ( \omega_{0} ~t ) ~-~
       \omega_{0} ~ \eta ~ \cos ( \omega_{0} ~t ) ~,
\nonumber \\
\omega_{0} &=&	A ~-~ B  ~,
\end{eqnarray}
if also equations (following from Eqs. 9)
\begin{eqnarray}\label{32}
\xi' &=& \xi~ \cos ( \omega_{0} ~t ) ~-~
	 \eta~ \sin ( \omega_{0} ~t ) ~,
\nonumber \\
\eta' &=& \xi~ \sin ( \omega_{0} ~t ) ~+~
	  \eta~ \cos ( \omega_{0} ~t ) ~,
\nonumber \\
\zeta ' &=& \zeta
\end{eqnarray}
and equations
\begin{eqnarray}\label{33}
\frac{d \xi '}{dt} &=&
       \frac{d \xi}{dt} ~ \cos ( \omega_{0} ~t ) ~-~
       \frac{d \eta}{dt} ~ \sin ( \omega_{0} ~t )
\nonumber \\
& & -~ \omega_{0} ~ \xi ~ \sin ( \omega_{0} ~t ) ~-~
       \omega_{0} ~ \eta ~ \cos ( \omega_{0} ~t ) ~,
\nonumber \\
\frac{d \eta '}{dt} &=&
       \frac{d \xi}{dt} ~ \sin ( \omega_{0} ~t ) ~+~
       \frac{d \eta}{dt} ~ \cos ( \omega_{0} ~t )
\nonumber \\
& & +~ \omega_{0} ~ \xi ~ \cos ( \omega_{0} ~t ) ~-~
       \omega_{0} ~ \eta ~ \sin ( \omega_{0} ~t ) ~,
\nonumber \\
\frac{d \zeta '}{dt} &=& \frac{d \zeta}{dt}
\end{eqnarray}
are used.

Eqs. (31) show that $\Gamma_{1}$ $\ne$ 0,
$\Gamma_{2}$ $\ne$ 0, $Z_{0}$ $\ne$ 0 violate
the time conservation of the quantity in the curly brackets
in the left-hand side of Eq. (31).

\section{Galactic tide for Dauphole et al. (1996) model of Galaxy}
Dauphole et al. (1996) model of the Galaxy yields
[see also Eqs. (29) in Kla\v{c}ka 2009]:
\begin{eqnarray}\label{34}
\frac{d^{2} \xi}{dt^{2}} &=& - ~ \frac{G M_{\odot}}{r^{3}} ~ \xi
~+~ ( A - B ) ~\left [ A + B + 2 A \cos \left ( 2 ~ \omega_{0} t \right )
 \right ] ~ \xi
\nonumber \\
& & -~ 2~ A~ ( A - B )~ \sin \left ( 2 ~ \omega_{0} t \right ) ~\eta
\nonumber \\
& & +~ ( A - B )^{2} ~( \Gamma_{1D} / \sqrt{b_{d}^{2} + Z_{0}^{2}}
 ~+~ \Gamma_{2D} ) ~R_{0} ~Z_{0}
~\cos \left ( \omega_{0} t \right ) ~\zeta ~,
\nonumber \\
\frac{d^{2} \eta}{dt^{2}} &=& - ~ \frac{G M_{\odot}}{r^{3}} ~ \eta
~-~  2 ~A~ ( A - B )~ \sin \left ( 2 ~ \omega_{0} t \right ) ~ \xi
\nonumber \\
& & +~ ( A - B )~ \left [ A ~+~ B ~-~ 2 ~A ~\cos \left ( 2 ~ \omega_{0} t \right )
 \right ] ~ \eta
\nonumber \\
& & -~ ( A - B )^{2} ~( \Gamma_{1D}  / \sqrt{b_{d}^{2} + Z_{0}^{2}}
~+~ \Gamma_{2D} ) ~R_{0} ~Z_{0}
~\sin \left ( \omega_{0} t \right ) ~\zeta ~,
\nonumber \\
\frac{d^{2} \zeta}{dt^{2}} &=& - ~ \frac{G M_{\odot}}{r^{3}} ~ \zeta
~-~ \left [ 4 ~\pi ~G ~\varrho ~+~
2 \left ( A^{2} ~-~ B^{2} \right ) \right ] ~\zeta
\nonumber \\
& & -~ 4 ~\pi ~G ~\varrho' ~
    Z_{0} \left [ \cos \left ( \omega_{0} t \right ) ~ \xi ~-~
    \sin \left ( \omega_{0} t \right ) ~ \eta \right ] ~,
\nonumber \\
\frac{d^{2} Z_{0}}{dt^{2}} &=& -~ \left [ 4 ~\pi ~G ~\varrho ~+~
2 \left ( A^{2} ~-~ B^{2} \right ) \right ] ~Z_{0} ~,
\nonumber \\
r &=& \sqrt{\xi ^{2} ~+~ \eta ^{2} ~+~ \zeta ^{2}} ~,
\nonumber \\
\omega_{0} &=& A ~-~ B ~,
\nonumber \\
A &=& 14.25 ~ km ~s^{-1} ~kpc^{-1} ~,
\nonumber \\
B &=&  -~ 13.89 ~ km ~s^{-1} ~kpc^{-1} ~,
\nonumber \\
\Gamma_{1D} &=& 0.084 ~kpc^{-1} ~,
\nonumber \\
\Gamma_{2D} &=& 0.008 ~kpc^{-2} ~,
\nonumber \\
\varrho &=& 0.143 ~M_{\odot}~ pc^{-3} ~,
\nonumber \\
\varrho' &=& -~ 0.0425 ~M_{\odot} ~pc^{-3} ~kpc^{-1} ~,
\nonumber \\
b_{d} &=& 0.25 ~kpc ~.
\end{eqnarray}

Equations analogous to Eqs. (21) are:
\begin{eqnarray}\label{35}
\frac{d^{2} \xi'}{dt^{2}} &=& - ~ \frac{G M_{\odot}}{r^{3}} ~\xi'
			      ~+~ 4~ A~ ( A~-~ B ) ~ \xi'
\nonumber \\
& & +~ ( A - B )^{2} ~( \Gamma_{1D} / \sqrt{b_{d}^{2} + Z_{0}^{2}}
 ~+~ \Gamma_{2D} ) ~R_{0} ~Z_{0} ~\zeta '
~-~ 2~ ( A~-~ B ) ~ \frac{d\eta'}{dt}  ~,
\nonumber \\
\frac{d^{2} \eta'}{dt^{2}} &=& - ~ \frac{G M_{\odot}}{r^{3}} ~\eta'  ~+~
 2~ ( A~-~ B ) ~ \frac{d\xi'}{dt} ~,
\nonumber \\
\frac{d^{2} \zeta'}{dt^{2}} &=& - ~ \frac{G M_{\odot}}{r^{3}} ~ \zeta '
~-~ \left [ 4 ~\pi ~G ~\varrho ~+~
2~ \left ( A^{2} ~-~ B^{2} \right ) \right ] ~\zeta '
\nonumber \\
& & -~ 4 ~\pi ~G ~\varrho' ~ Z_{0}  ~ \xi' ~,
\nonumber \\
\frac{d^{2} Z_{0}}{dt^{2}} &=& -~ \left [ 4 ~\pi ~G ~\varrho ~+~
2 ~\left ( A^{2} ~-~ B^{2} \right ) \right ] ~Z_{0} ~,
\nonumber \\
r &=& \sqrt{\xi^{'~2} ~+~ \eta^{'~2} ~+~ \zeta^{'~2}} ~,
\nonumber \\
\omega_{0} &=& A ~-~ B ~.
\end{eqnarray}

The vector of the perturbing acceleration acting on the comet is
\begin{eqnarray}\label{36}
\vec{F}' &=& ( \vec{F}' )_{\xi '} ~\hat{\vec{\xi}} ' ~+~
( \vec{F}' )_{\eta '} ~\hat{\vec{\eta}} ' ~+~
( \vec{F}' )_{\zeta '} ~\hat{\vec{\zeta}} ' ~,
\nonumber \\
( \vec{F}' )_{\xi '} &=& ( A ~-~ B ) ~\left \{ 4~ A~  \xi'
~+~ ( A - B ) ~\left ( \frac{\Gamma_{1D}}{\sqrt{b_{d}^{2} + Z_{0}^{2}}}
 ~+~ \Gamma_{2D} \right ) ~R_{0} ~Z_{0} ~\zeta '
~-~ 2~ \frac{d\eta'}{dt} \right \}  ~,
\nonumber \\
( \vec{F}' )_{\eta '}  &=&  2~ ( A ~-~ B ) ~ \frac{d\xi'}{dt} ~,
\nonumber \\
( \vec{F}' )_{\zeta '} &=& -~ \left [ 4 ~\pi ~G ~\varrho ~+~
2~ \left ( A^{2} ~-~ B^{2} \right ) \right ] ~\zeta' ~-~
4 ~\pi ~G ~\varrho' ~ Z_{0}  ~ \xi' ~.
\end{eqnarray}

Finally, equations analogous to Eqs. (31) are:
\begin{eqnarray}\label{37}
LHS_{S~D} &=& RHS_{S~D} ~,
\nonumber \\
LHS_{S~D} &=& \frac{d}{dt} ~\left \{ -~\frac{G~M_{\odot}}{2 ~p} ~\left (
1 ~-~ e^{2} \right ) ~+~ ( A - B ) ~ \sqrt{G ~M_{\odot}~p} ~\cos i
\right .
\nonumber \\
& & \left .  + ~\frac{1}{2} ~ ( A ~-~ B )^{2} ~ \left ( \xi ^{2} ~+~
\eta ^{2} \right ) \right .
\nonumber \\
& & \left .  -~ 2 ~A ~ ( A - B ) ~ [ \xi~ \cos ( \omega_{0} ~t ) ~-~
	     \eta~ \sin ( \omega_{0} ~t )  ] ^{2} \right .
\nonumber \\
& & \left . +~ ( 2 ~\pi ~G ~\varrho ~+~
A^{2} ~-~ B^{2} ) ~ \zeta ^{2} \right \} ~,
\nonumber \\
RHS_{S~D} &=& ( A - B )^{2} ~( \Gamma_{1D} / \sqrt{b_{d}^{2} + Z_{0}^{2}}
 ~+~ \Gamma_{2D} ) ~R_{0} ~Z_{0} ~\zeta ~ \times V
\nonumber \\
& & -~ 4 ~\pi ~G ~\varrho' ~ Z_{0}  ~
[ \xi~ \cos ( \omega_{0} ~t ) ~-~ \eta~ \sin ( \omega_{0} ~t ) ]
~\frac{d \zeta}{dt}  ~,
\nonumber \\
V &=& \frac{d \xi}{dt} ~ \cos ( \omega_{0} ~t ) ~-~
       \frac{d \eta}{dt} ~ \sin ( \omega_{0} ~t )
\nonumber \\
& & -~ \omega_{0} ~ \xi ~ \sin ( \omega_{0} ~t ) ~-~
       \omega_{0} ~ \eta ~ \cos ( \omega_{0} ~t ) ~,
\nonumber \\
\omega_{0} &=& A - B ~.
\end{eqnarray}

\section{Conclusion}
The paper treats the effect of the galactic tide on motion of a comet with
respect to the Sun. Eqs. (21) correspond to the relevant equation of motion
in the rotating noninertial frame of reference (the $\xi' -$axis is still
oriented outward from the galactic center, the $\zeta '-$axis is oriented toward
the north Galactic pole and the $\xi' -$, $\eta ' -$ and $\zeta ' -$ axes form
the right-handed coordinate system). The vector of the perturbing force acting
on the comet is given by Eqs. (22). The equation of motion  produces Eq. (31)
which would represent a time independent quantity for the conventionally
considered form of the galactic tidal field characterized by the conditions
$\Gamma_{1}$ $=$ 0, $\Gamma_{2}$ $=$ 0, $Z_{0}$ $\equiv$ 0 (in reality,
the first two conditions or the third condition alone are sufficient).

Solving Eqs. (21) for a comet, one can find an orbital evolution of the comet
in terms of orbital elements. System of Eqs. (21), (23)-(24) and (26) has to
be used. Sec. 4.1 describes how to transform the solution
of Eqs. (21) into the evolution of the osculating orbital elements.

The case of Dauphole et al. model of Galaxy is presented by Eqs. (34)-(37).

\section*{Appendix A: Inertial and noninertial frames of reference}

(Reference to equation of number (j) of this appendix is denoted as Eq. (A~j).
Reference to equation of number (i) of the main text is denoted as Eq. (i).)

\setcounter{equation}{0}

The text presented between Eqs. (10) and (11) defines the velocity
in the noninertial frame of reference
(the velocity with respect to the system $S'$) in the form $\vec{v}'$ $\equiv$
$d' \vec{r} ' / dt$ $\equiv$ ($dx'/dt$) $\hat{\vec{x}}'$ $+$
($dy'/dt$) $\hat{\vec{y}}'$ $+$ ($dz'/dt$) $\hat{\vec{z}}'$.
The question "Why is there a prime above the symbol of differentiation
in $d' \vec{r} ' / dt$?" may appear. This appendix explicitly explains
the situation.

At first, we have position vectors:
\begin{eqnarray}\label{A1}
\vec{r}' &=& \vec{r} ~,
\end{eqnarray}
\begin{eqnarray}\label{A2}
\vec{r}' &=& x'~\hat{\vec{x}} ' ~+~ y'~\hat{\vec{y}} ' ~+~ z'~\hat{\vec{z}} ' ~,
\end{eqnarray}
\begin{eqnarray}\label{A3}
\vec{r} &=& x~\hat{\vec{x}} ~+~ y~\hat{\vec{y}} ~+~ z~\hat{\vec{z}}  ~.
\end{eqnarray}

As for the velocity vector, we have
\begin{eqnarray}\label{A4}
\vec{v} &=& \frac{d \vec{r}}{dt} ~,
\nonumber \\
\vec{v} &=& \frac{d x}{dt} ~\hat{\vec{x}} ~+~ \frac{d y}{dt} ~\hat{\vec{y}}
	    ~+~ \frac{d z}{dt} ~\hat{\vec{z}}
\end{eqnarray}
in the inertial frame of reference, if Eq. (A3) is used.
We can also write, on the basis of Eqs. (A1) and (A2),
\begin{eqnarray}\label{A5}
\vec{v} &=& \frac{d \vec{r}}{dt} = \frac{d \vec{r}'}{dt}
\nonumber \\
&=& \frac{d }{dt} \left (
    x'~\hat{\vec{x}} ' ~+~ y'~\hat{\vec{y}} ' ~+~ z'~\hat{\vec{z}} ' \right )
\nonumber \\
&=& \frac{d x'}{dt} ~\hat{\vec{x}}' ~+~ x' ~\frac{d \hat{\vec{x}}'}{ dt} ~+
\nonumber \\
& & \frac{d y'}{dt} ~\hat{\vec{y}}' ~+~ y' ~\frac{d \hat{\vec{y}}'}{ dt} ~+
\nonumber \\
& & \frac{d z'}{dt} ~\hat{\vec{z}}' ~+~ z' ~\frac{d \hat{\vec{z}}'}{ dt} ~.
\end{eqnarray}
The last relation can be arranged on the basis of Eqs. (7):
\begin{eqnarray}\label{A6}
\vec{v} &=& \frac{d x'}{dt} ~\hat{\vec{x}}' ~+~ x' ~\left \{
-~ \omega ~ \sin \left ( \omega ~t \right ) ~\hat{\vec{x}} ~+~
   \omega ~ \cos \left ( \omega ~t \right ) ~\hat{\vec{y}} \right \} ~+~
\nonumber \\
& & \frac{d y'}{dt} ~\hat{\vec{y}}' ~+~ y' ~\left \{
-~ \omega ~ \cos \left ( \omega ~t \right ) ~\hat{\vec{x}} ~-~
   \omega ~ \sin \left ( \omega ~t \right ) ~\hat{\vec{y}} \right \} ~+~
\nonumber \\
& & \frac{d z'}{dt} ~\hat{\vec{z}}'
\nonumber \\
&=& \frac{d x'}{dt} ~\hat{\vec{x}}' ~+~ \frac{d y'}{dt} ~\hat{\vec{y}}' ~+~
    \frac{d z'}{dt} ~\hat{\vec{z}}' ~+~
\nonumber \\
& & -~ \omega ~\left \{ x' ~ \sin \left ( \omega ~t \right ) ~+~ y' ~
    \cos \left ( \omega ~t \right ) \right \} ~\hat{\vec{x}}
\nonumber \\
& & +~ \omega ~\left \{ x' ~ \cos \left ( \omega ~t \right ) ~-~ y' ~
    \sin \left ( \omega ~t \right ) \right \} ~\hat{\vec{y}} ~.
\end{eqnarray}
Using Eqs. (10), we obtain
\begin{eqnarray}\label{A7}
\vec{v} &=& \vec{v}' ~-~ \omega ~ y ~\hat{\vec{x}} ~+~ \omega ~x
	    ~\hat{\vec{y}}
\nonumber \\
&=& \vec{v}' ~+~ \vec{\omega} \times \vec{r}' ~,
\end{eqnarray}
since $\vec{v}'$ $=$ ($dx'/dt$) $\hat{\vec{x}}'$ $+$
($dy'/dt$) $\hat{\vec{y}}'$ $+$ ($dz'/dt$) $\hat{\vec{z}}'$ $\equiv$
$d' \vec{r} ' / dt$.

The last relation of Eq. (A5) can be arranged in an another way.
We can use
\begin{eqnarray}\label{A8}
\frac{d \hat{\vec{x}}'}{ dt} &=&
\vec{\omega} ~\times ~\hat{\vec{x}}' ~,
\nonumber \\
\frac{d \hat{\vec{y}}'}{ dt} &=&
\vec{\omega} ~\times ~\hat{\vec{y}}' ~.
\end{eqnarray}
Eqs. (A5) and (A8) immediately yield
\begin{eqnarray}\label{A9}
\vec{v} &=&
\frac{d x'}{dt} ~\hat{\vec{x}}' ~+~ x' ~
\vec{\omega} ~\times ~\hat{\vec{x}}' ~+~
\nonumber \\
& & \frac{d y'}{dt} ~\hat{\vec{y}}' ~+~ y' ~
\vec{\omega} ~\times ~\hat{\vec{y}}' ~+~
\nonumber \\
& & \frac{d z'}{dt} ~\hat{\vec{z}}'
\nonumber \\
&=& \vec{v}' ~+~ \vec{\omega} \times \vec{r}' ~.
\end{eqnarray}
The final results of Eqs. (A7) and (A9) are equal.

As for acceleration, the results represented by Eqs. (A7) and (A9)
yield
\begin{eqnarray}\label{A10}
\frac{d \vec{v}}{dt} &=& \frac{d }{dt} \left (
\vec{v}' ~+~ \vec{\omega} \times \vec{r}' \right )
\nonumber \\
&=& \frac{d \vec{v}'}{dt} ~+~ \frac{d }{dt} \left (
\vec{\omega} \times \vec{r}' \right )
\nonumber \\
&=& \frac{d' \vec{v}'}{dt} ~+~ \vec{\omega} \times \vec{v}' ~+~
\vec{\omega} \times \frac{d \vec{r}'}{dt}
\nonumber \\
&=& \frac{d' \vec{v}'}{dt} ~+~ \vec{\omega} \times \vec{v}' ~+~
\vec{\omega} \times \left (
\vec{v}' ~+~ \vec{\omega} \times \vec{r}' \right ) ~,
\end{eqnarray}
or, simply,
\begin{eqnarray}\label{A11}
\vec{a} &=& \vec{a}' ~+~ 2~ \vec{\omega} \times \vec{v}' ~+~
\vec{\omega} \times \left (
\vec{\omega} \times \vec{r}' \right ) ~.
\end{eqnarray}

\section*{Acknowledgement}
This work was supported by the Scientific Grant Agency VEGA, Slovak Republic,
grant No. 2/0016/09.

\end{document}